\documentstyle[twoside]{article}
\newcommand{\scatola}[1]{\Blue\fbox{$\displaystyle #1$}\Black}
\newcount\Mac  \Mac=0  
\newcommand{\ifMac}[2]{\ifnum\Mac=1 #1 \else #2 \fi}
\newcommand{\riga}[1]{\noalign{\hbox{\parbox{\textwidth}{#1}}}\nonumber}
\newcommand{\GF}{G_{\rm F}}

\newcommand{\One}{\hbox{1\kern-.24em I}}

\newcommand{\BR}{\mathop{\rm BR}}
\newcommand{\MSbar}{{\overline{\hbox{\sc ms}}}}

\newcommand{\Dsl}{D\hspace{-6.6pt}{/}\,}
\newcommand{\Wsl}{W\hspace{-9pt}{/}\,\,}

\newcommand{\ds}{\partial\!\!\!\raisebox{2pt}[0pt][0pt]{$\scriptstyle/$}\,}
\newcommand{\GeV}{\,{\rm GeV}}

\newcommand{\NP}{Nucl. Phys.}
\newcommand{\PRL}{Phys. Rev. Lett.}
\newcommand{\PL}{Phys. Lett.}
\newcommand{\PR}{Phys. Rev.}
\newcommand{\pL}{{\cal P}_{\rm L}} 
\newcommand{\pR}{{\cal P}_{\rm R}}

\newcommand{\eq}[1]{~{\rm (\ref{eq:#1})}}

\newcommand{\lnEps}{\ln\frac{\mub^2}{m_t^2}}
\newcommand{\epsIR}{\varepsilon_{\rm ir}}
\newcommand{\epsUV}{\varepsilon_{\rm uv}}
\newcommand{\eps}{\varepsilon}
\newcommand{\mub}{\bar{\mu}}
\newcommand\Ord{{\cal O}}
\newcommand\Op[1]{{\cal O}_{#1}}
\newcommand{\pl}{p\hspace{-4.2pt}{\scriptstyle/}}

\newcommand{\Li}{\hbox{Li}_2}

\def\Red{}
\def\Black{}
\def\Blue{}

\newcommand{\lascia}[1]{}
\makeatletter
\def\art{\@ifnextchar[{\eart}{\oart}}
\def\eart[#1]#2#3#4#5#6{{\rm #2}, {\rm #3 \bf #4} {\rm (#6) #5} ({\em #1})}
\def\hepart[#1]#2{{\rm #2, \em#1}}
\newcommand{\oart}[5]{{\rm #1}, {\rm #2 \bf #3} {\rm (#5) #4}}
\newcommand{\y}{{\rm and} }
\newcounter{alphaequation}[equation]
\def\thealphaequation{\theequation\hbox to
0.6em{\hfil\alph{alphaequation}\hfil}}
\def\eqnsystem#1{
\def\@eqnnum{{\rm (\thealphaequation)}}
\def\@@eqncr{\let\@tempa\relax \ifcase\@eqcnt \def\@tempa{& & &} \or
  \def\@tempa{& &}\or \def\@tempa{&}\fi\@tempa
  \if@eqnsw\@eqnnum\refstepcounter{alphaequation}\fi
\global\@eqnswtrue\global\@eqcnt=0\cr}
\refstepcounter{equation} \let\@currentlabel\theequation \def\@tempb{#1}
\ifx\@tempb\empty\else\label{#1}\fi
\refstepcounter{alphaequation}
\let\@currentlabel\thealphaequation
\global\@eqnswtrue\global\@eqcnt=0 \tabskip\@centering\let\\=\@eqncr
$$\halign to \displaywidth\bgroup \@eqnsel\hskip\@centering
$\displaystyle\tabskip\z@{##}$&\global\@eqcnt\@ne
\hskip2\arraycolsep\hfil${##}$\hfil& \global\@eqcnt\tw@\hskip2\arraycolsep
$\displaystyle\tabskip\z@{##}$\hfil
\tabskip\@centering&\llap{##}\tabskip\z@\cr}

\def\endeqnsystem{\@@eqncr\egroup$$\global\@ignoretrue} \makeatother

\def\Ord{{\cal O}}

\def\SU{{\rm SU}}

\def\circa#1{\,\raise.3ex\hbox{$#1$\kern-.75em\lower1ex\hbox{$\sim$}}\,}

\begin{document}
\begin{quote}
\makebox[0cm][l]{April 1998} \hfill 
\makebox[0cm]{IFUP--TH 16/98}\hfill
\makebox[0cm][r]{hep-ph/9804274}
\end{quote}
\bigskip\bigskip
\centerline{\LARGE\bf\Red Two-loop heavy top corrections to the $b\to s \gamma$ decay}
\bigskip\bigskip\Black
\centerline{\large\bf Alessandro Strumia} \vspace{0.3cm}
\centerline{\em Dipartimento di Fisica, Universit\`a di Pisa and}
\centerline{\em INFN, sezione di Pisa,  I-56126 Pisa, Italia}\vspace{0.3cm}
\bigskip\bigskip\Blue
\centerline{\large\bf Abstract}
\begin{quote}\large\indent
We compute the two-loop corrections to the coefficient
of the $b\to s\gamma$ magnetic penguin present in the limits of
heavy top and/or heavy higgs.
This kind of corrections affects in a significant way the observables measured at LEP.
On the contrary we find that, due to a numerical accident,
the correction to $\BR(B\to X_s\gamma)$ is negligible (below the $1\%$ level
for any possible value of the higgs mass)
when the leading order result is expressed in the usual way
in terms of the semileptonic $\BR(B\to X_c \bar{e}\nu)$.

\end{quote}\Black

\section{Introduction}
The $b\to s\gamma$ decay seems the more promising observable in $B$, $K$ and $D$ physics
to search for a supersymmetric effect.
Some observables (like the $K\to \pi \nu\bar{\nu}$ decays and various tree level processes)
are predictable with remarkable accuracy within the SM~\cite{B},
but are expected to receive too small supersymmetric corrections.
In other cases (like in $B$ mixing, or in charm-less $B$ decays)
the supersymmetric corrections can be large enough to affect the experimental result,
but remain masked by too large hadronic uncertainties in the SM prediction~\cite{generalSUSYinB,hadronic}.
Even if the minimal supersymmetric SM (MSSM) Lagrangian contains
the new sources of flavour and CP violation that can be left at low energy
by unification-scale physics~\cite{LFV},
it does not seem possible to disentangle the new supersymmetric effects in $B$-physics
from the SM background~\cite{SUSYinB}.
More optimistic conclusions can only be obtained assigning ad hoc values to the
$\sim$ 100 parameters of the MSSM~\cite{adhocSUSYinB},
or if $R$-parity is broken in appropriate way.
Various supersymmetric effects become larger if a stop state is light
(this possibility allows baryogenesis at the electroweak scale~\cite{EWbaryogenesis},
but requires an unnaturally large fine tuning~\cite{lightStop})
or if $\tan\beta$ is large (this possibility can be realized naturally with some appropriate
hierarchy between the parameters of the higgs potential).
In both cases also the supersymmetric correction to $b\to s\gamma$ is enhanced.

In all cases, it is interesting to study the $b\to s\gamma$ magnetic penguin and
it is reasonable to expect that its Wilson coefficient, usually named $C_7$, receives
a relatively large and hopefully detectable $\Ord(10\%)$ supersymmetric correction.
Consequently, it is useful to perform accurate multi-loop computations of this observable.

In order to obtain a sufficiently accurate SM prediction, the next-to-leading (NLO)
{\em QCD corrections\/},
of relative order $g_3^2(\mub)/4\pi$, have been computed
for all relevant values of the $\MSbar$ scale $\mub$:
for $\mub$ near the electroweak scale $\mub_W$~\cite{MWmatch},
for $\mub$ near the $B$         scale $\mub_b$~\cite{MBmatch},
and for $\mub$ between $\mub_W$ and $\mub_b$~\cite{NLO7}.
All these three stages give numerically relevant
QCD corrections of order 20\% to the decay rate.
Also the supersymmetric contributions to $C_7$ are affected by relevant QCD corrections.
However, only the QCD corrections to the charged higgs mediated contribution are known~\cite{C71H}.
It is difficult to present in a compact and accurate form the SUSY-QCD correction to the
remaining relevant supersymmetric contribution, mediated by charginos and squarks.

{\em Electroweak corrections\/},
of relative order $g_2^2/(4\pi)^2$ and $\ln (\mub_W^2/\mub_b^2)\cdot e^2/(4\pi)^2$,
have partly been computed in~\cite{CzMa} and found to be relevant.

\smallskip

In this paper we compute, for arbitrary value of the Higgs mass,
the {\em heavy top corrections\/} of relative order $g_t^2/4\pi$
where $g_t$ is the top quark Yukawa coupling in the SM.
We also compute the corrections of relative order $\lambda^2/4\pi$
where $\lambda$ is the quartic higgs coupling in the SM.
More precisely the couplings $g_t$ and $\lambda$ induce corrections that
depend on the top, higgs and $W$ boson squared masses $(m_t^2,m_h^2,m_W^2)$,
and increase when $m_t$ or $m_h$ become large,
as summarised in table~\ref{tab:lims}.
We compute these potentially relevant corrections in the limit $m_W\to 0$.
In the same approximation, the `heavy top corrections' and/or the
`heavy higgs corrections' 
to the precision electroweak observables measured at LEP
have been computed in~\cite{mt^4} and/or~\cite{lnmh}.
The heavy top corrections to LEP observables
are so relevant that also the sub-leading terms,
suppressed by a power of $m_W^2/m_t^2$, have recently been
computed~\cite{Degrassi}.
From these results it is possible to derive the heavy top corrections to
various $B$ and $K$ decays~\cite{Kpinu} generated by `electric' effective operators.
On the contrary, a new computation is necessary to obtain
the corrections to the $b\to s\gamma$ magnetic penguin.

This computation is outlined in section~2 (various details are confined to the appendices) and
is more cumbersome than in the other mentioned cases.
It is not possible to relate the magnetic penguin to a simpler vertex, as in~\cite{mt^4}.
As a consequence more than 50 two loop diagrams, shown in fig.~\ref{fig:gra2}, need to be evaluated.
Renormalization is a complicated task:
we cannot set the $b$-quark mass $m_b$ to zero,
we have to deal with CKM mixing,
and we have to connect $b\to s\gamma$ with the measurable $B\to X_s\gamma$.
A particularly appropriate way of doing the renormalization is described in sections~2.3 and 2.4.
The final result is discussed in section~3.
We find that {\em the percentage correction to $\BR(B\to X_s\gamma)/\BR(B\to X_c\bar{e}\nu)$ is negligible\/} ---
at the $1\%$ level --- for any interesting value of the higgs mass.
This is due to a numerical accident; the single contributions would separately give few $\%$ corrections.
The presence of accidental cancellations suggests that the approximation $m_W\to 0$
could be not a good one.
In any case it appears unlikely that the neglected corrections,
suppressed by powers of $m^2_W/m^2_t$, be sufficiently large to
make the heavy top and/or heavy higgs effects relevant.

\section{Computation}
In this section we outline the computation.
For simplicity we will refer to the heavy top and/or heavy higgs corrections that
we want to compute as `$\Ord(g_t,\lambda)^2$ corrections',
even if this is not exact.
The relevant Lagrangian is given in appendix~A, where the various parameters are defined.

\subsection{One loop result}
We begin with recalling the structure of the one-loop result.
The contribution of the first generation is negligible,
so that the unitarity constraint on the CKM matrix becomes $V_{cb} V_{cs}^* +V_{tb} V_{ts}^*=0$.
It is also unnecessary (even it it would be immediate)
to compute the $\Ord(g_t,\lambda)^2$ corrections to
terms suppressed by the ratio $m_s/m_b$ between the masses of the strange and the bottom quarks.
The effective Hamiltonian for the $b\to s\gamma$ decay is
\begin{equation}\label{eq:Heff}
{\cal H}_{\rm eff} = -\frac{\GF}{\sqrt{2}}V_{tb} V_{ts}^*\sum_{i=1}^8
C_i {\cal O}_i
\end{equation}
where $\GF$ is the Fermi constant, $C_i$ are the Wilson coefficients and,
for our purposes, the relevant operators are
$$
{\cal O}_2 \equiv 4(\bar{s}^i\gamma_\mu \pL c^i)(\bar{c}^j\gamma_\mu \pL b^j) ,\qquad
{\cal O}_7 \equiv 4\frac{e}  {(4\pi)^2}m_b(\bar{s}_i\gamma_{\mu\nu}          \pR b_i)  F_{\mu\nu},\qquad
{\cal O}_8 \equiv 4\frac{g_3}{(4\pi)^2}m_b(\bar{s}_i\gamma_{\mu\nu}T^a_{ij}  \pR b_j)G^a_{\mu\nu}
$$
where $i,j$ are colour indexes.
At leading order, and before including QCD corrections, $C_2=1$.
The leading contribution to $C_7$ is given,
in the the renormalizable
Feynman-Fujikawa gauge~\cite{gauge}, by the one-loop graphs (named ``W'', ``I'' and ``E'')
shown in fig.~\ref{fig:gra1}:
\begin{equation}\label{eq:P}
C_7=\big[P_W(x_t)-P_W(x_c)\big]+\big[-\frac{x_t}{2} P_{I}(x_t)  + \frac{x_t}{2} P_{E}(x_t)\big]=
\frac{3}{2} x_t P_{E}(x_t)\approx -0.2
\end{equation}
where $x_t=m_t^2/m_W^2$, $x_c=m_c^2/m_W^2\approx 0$ and
the three penguin loop functions $P(x)$ are given in appendix~C.
In the limit $m_t\gg m_W$ (in which we will compute the $\Ord(g_t,\lambda)^2$ corrections)
$P_W(x_t\to\infty)=0$ and the one-loop result\eq{P} simplifies to
\begin{equation}
C_7(m_t\gg m_W)=\big[0-\frac{23}{36}\big]  +  \big[\frac{5}{12}-\frac{1}{9}\big]  =  -\frac{1}{3}.
\end{equation}

\subsection{Two loop $\Ord(g_t,\lambda)^2$ graphs}
\begin{figure}[t]\setlength{\unitlength}{1in}\begin{center}
\begin{picture}(6,0.7)
\ifMac
{\put(0,-0.3){\special{picture gra1}}}
{\put(0,-0.3){\includegraphics{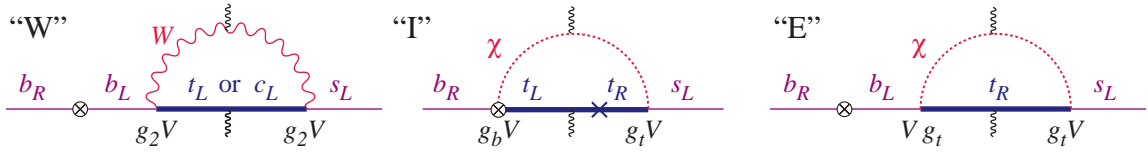}}}
\end{picture}
\vspace{0.5cm}
\caption[SP]{\em The one-loop $b\to s\gamma$ graphs.
The symbol $\otimes$ denotes a $m_b$ factor.
All possible attachments of the photon have been shown.
\label{fig:gra1}}
\end{center}\end{figure}

We have seen that the one-loop $b\to s\gamma$ decay, in the limit of heavy top, does not receive any contribution
from the graphs ``W'' of fig.~\ref{fig:gra1}
with a $W$-boson and a top quark in the loop.
It is easy to understand this fact with a simple dimensional analysis and a look at the vertices,
explicitly written in fig.~\ref{fig:gra1}.
For the same reason~\cite{mt^4} {\em also at two loops the heavy top and/or heavy higgs corrections
can be computed in the gauge-less limit}.
In the same way the contribution mediated by the $W$ boson and the charm quark does not receive heavy top and/or heavy higgs corrections.
At the light of these considerations, all the two-loops graphs that give $\Ord(g_t,\lambda)^2$ corrections
are shown in fig.~\ref{fig:gra2}
(the 12 graphs named `B', `$h$' and `$\chi_L$' vanish in the limit of zero higgs mass).
We have not plotted other 10 graphs of tadpole type
(we will discuss them in the following).

All the graphs contain a (bottom quark $b$)-(top quark $t$)-(unphysical charged higgs $\chi$) vertex.
The SM Lagrangian\eq{Lclass} contains two different vertices of this kind:
$$g_b ~ \bar{b}_R t_L\chi^-\qquad\hbox{and}\qquad -g_t ~ \bar{b}_L t_R~\chi^-.$$
With the exception of the `$h_L$', `$\eta_L$' and `$\chi_L$' graphs
(marked with a $\otimes$ in fig.~\ref{fig:gra2})
both vertices contribute to the $b_R\to s_L \gamma$ magnetic penguin,
and give two different loop structures.
The difference is clear from the one-loop graphs of fig.~\ref{fig:gra1}, where we have separately plotted
the two contributions, named ``I'' and ``E''.
The ``E''-type graphs are usually more cumbersome to compute,
give smaller contributions
and generate a different $b\to s\gamma$ operator, equivalent to ${\cal O}_7$ only on-shell.

In total it is necessary to compute 64 two loop graphs.
This computation can be done in few minutes using a {\tt Mathematica}~\cite{Math} code.
Since we never need to define traces
like $\hbox{Tr}\,\gamma_5\gamma_\mu\cdots \gamma_\rho$,
nor the completely antisymmetric tensor,
we can employ na\"{\i}ve dimensional regularization
(i.e.\ anticommuting $\gamma_5$)
with $\overline{\hbox{MS}}$ renormalization scale~$\mub$.

\subsection{Renormalization procedure}
Unrenormalized tree level quantities will be denoted with a supscript $^0$.
The renormalization factor of a field $\phi$ ($\phi=\{t_L,t_R,b_L,b_R,\chi,\ldots\}$) is named $Z_\phi$,
and $\phi^0=\phi/\sqrt{Z_\phi}$.
The renormalization factor of a parameter $\wp$ ($\wp=\{m_t,m_b,v,V_{ts},\ldots\})$ is named $Z_\wp$,
and $Z_\wp \wp^0$ is $\wp^0$ plus the quantum corrections to it.
All the renormalization $Z$ factors are defined in a more precise way in appendix~B.

\begin{table}[p]
$$\begin{array}{|c|cc|cc|}\hline
\hbox{asymptotic dependence}&
\multicolumn{2}{|c|}{\hbox{\Blue large $m_t$ limit\Black}}&
\multicolumn{2}{|c|}{\hbox{\Blue large $m_h$ limit\Black}}\\
\hbox{of the corrections to}&\hbox{one loop} & \hbox{two loops}&\hbox{one loop} & \hbox{two loops}\\ \hline
\Blue\hbox{SM electroweak observables}\Black & m_t^2 &m_t^4&\ln m_h&m_h^2\\
\Blue b\to s\gamma\hbox{ magnetic penguin}\Black & m_t^0 &m_t^2&{\rm Cte}&\ln m_h\\ \hline
\end{array}$$\Black
\caption[24]{\em Asymptotic dependence of the heavy top and of the heavy higgs corrections
to a typical observable measured ad LEP, and to the coefficient of the
$b\to s\gamma$ magnetic penguin.\label{tab:lims}}
\end{table}

\begin{figure}[p]\setlength{\unitlength}{1in}\begin{center}
\begin{picture}(5,5.25)
\ifMac
{\put(0,-0.3){\special{picture gra2}}}
{\put(0,-0.3){\includegraphics{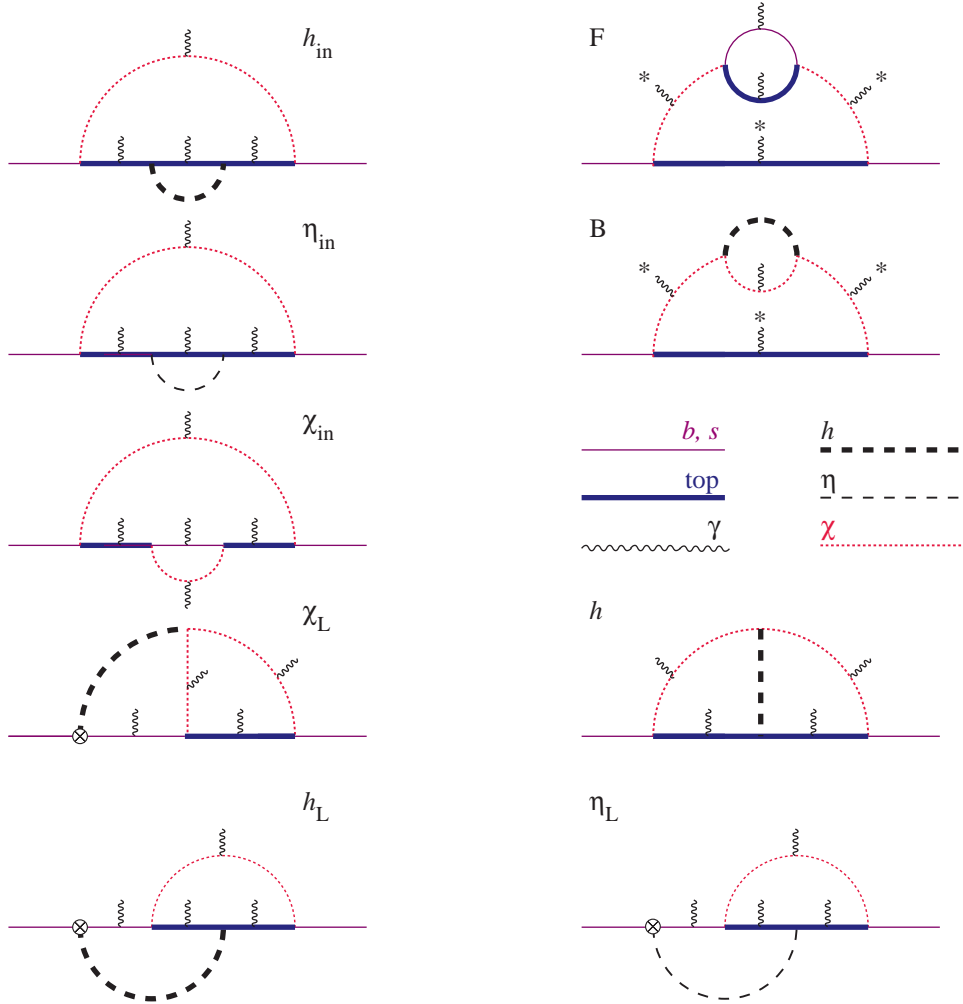}}}
\end{picture}
\vspace{0.5cm}
\caption[SP]{\em The two loop-diagrams that give $\Ord(g_t,\lambda)^2$ corrections.
The thick (thin) continuous lines represent the top ($b$ and $s$) quark.
The dotted (dashed) thin lines represent the charged (neutral) unphysical higgses $\chi$ ($\eta$).
The thick dashed line represent the SM higgs boson $h$.
The thin wavy lines represent all possible attachments of the photon.
In the graphs marked with a $\ast$ the correction on the $\chi$ propagator has to be subtracted
at zero $\chi$ momentum.
Tadpole diagrams are not shown.
\label{fig:gra2}}
\end{center}\end{figure}

Renormalization would be a {\em very\/} cumbersome task if done
adding counterterm diagrams to cancel the one-loop subdivergences.
It is more convenient to proceed in a different way.
Renormalize amounts to express the quantum-corrected $b\to s\gamma$ amplitude
in terms of quantum-corrected physical quantities:
the bare parameters must be substituted with measurable quantities,
and the $b,s,\gamma$ fields have to be re-normalized.
In the limit $m_t\gg m_W$ the one loop $b\to s\gamma$ amplitude depends on the following bare parameters:
$g_t^0$ (or $m_t^0$), $g_b^0$ (or $m_b^0$), $v^0$ (or $\GF^0$,
that appears as the overall dimensional factor in\eq{Heff}), the electric charge $e$
(that do not get any $\Ord(g_t,\lambda)^2$ correction, because it must be the same for any quark or lepton)
and the CKM mixing angle between the two heavy generations.
We define the renormalized parameters in terms of precisely measured quantities:
\begin{itemize}
\item the pole top mass, $m_t$, given by
\begin{equation}
m_t = m_t^0\frac{Z_{m_t}}{\sqrt{Z_{t_L} Z_{t_R}}}\qquad(m_t^0 = g_t^0 v^0)
\end{equation}
In the limit $m_t\gg m_W$ the one loop contribution to $C_7$ depends on $m_t$
only through terms that vanish as $\eps\to 0$, so that
only the divergent parts of the $Z$ give a contribution.
It is thus irrelevant to
distinguish between the top pole mass and top running mass parameter.

\item the pole bottom mass, $m_b$, given by
\begin{equation}
m_b = m_b^0\frac{Z_{m_b}}{\sqrt{Z_{b_L} Z_{b_R}}}\qquad(m_b^0 = g_b^0 v^0)
\end{equation}
Since the one-loop $\Ord(g_t^2)$ correction to the $b$ propagator
is mediated by a heavy top, these is no difference between pole (on-shell) and zero-momentum bottom mass.

\item  the Fermi constant measured in $\mu$ decay, $G_{\rm F}$.
Including the $\Ord(g_t,\lambda)^2$
 corrections, $\GF$ (or the $W$ mass term at zero momentum, $m_W$)
is linked to the higgs vacuum expectation value (vev) $v$ by~\cite{mt^4}
\begin{equation}\label{eq:MW<-->v}
\frac{4\GF}{\sqrt{2}}=\frac{g_2^2}{2m_W^2}=
\frac{1}{v^2 Z_{\chi}}
\end{equation}
where $Z_\chi$ is precisely defined in appendix B.3.
The relation\eq{MW<-->v}, valid in the gauge-less limit, can be obtained~\cite{mt^4}
considering the boson vectors as external currents and noticing that the
Ward identities of the $\SU(2)_L$ global symmetry
imply that the effective Lagrangian contains the following term
$${\cal L}_{\rm eff}=Z_\chi\left|\partial_\mu \chi-i\frac{g_2}{v}{\sqrt{2}} W_\mu\right|^2+\cdots=
m_W^2 |W|^2+\cdots$$

\item The higgs vev $v$ must be determined from the minimisation condition of the effective potential.
In the gauge-less limit this is equivalent to require that the unphysical higgses
$\chi$ and $\eta$ be massless Goldstone bosons:
\begin{equation}\label{eq:m=0}
m_\chi^2=\mu^2-2\lambda v^2+\hbox{quantum corrections of order }\frac{g_t^2m_t^2}{(4\pi)^2}
\hbox{ and smaller}=0
\end{equation}
At this point we encounter a technical problem: there are corrections to $v^2$ of order $g_t^4\,v^2$.
When $v_0$ is expressed in terms of $v$,
the term in the one loop coefficient of the $b\to s\gamma$ penguin suppressed by one power of $x_t^{-1}=m_W^2/m_t^2$,
$~C_7=-1/3+ 3(\ln x_t)/4x_t+\cdots$,
would give a heavy top correction, infrared (IR) divergent in the limit $m_W\to 0$,
that cancels a similar IR divergence present in the two loop graphs.
To avoid these infrared problems, it is more convenient to cancel the quantum corrections to the
Goldstone masses in eq.\eq{m=0} adding appropriate counterterms in order to obtain $v=v^0$.
This procedure requires to include the new $b\to s\gamma$ Feynman graphs that contain these counterterms.
This is equivalent to compute the two-loop graphs named ``F'' and ``B'' and marked with an $\ast$ in fig.~\ref{fig:gra2} with the prescription that
{\em the correction to the $\chi$ propagator
has to be subtracted at zero $\chi$ momentum}\footnote{For this reason we have not plotted the graphs
with a tadpole correction to the $\chi$ propagator:
they are cancelled by this procedure.
The tadpole graph in which the external photon is attached to the tadpole loop
also gives a zero contribution to the $b\to s\gamma$ magnetic penguin.}.
When this is done, all infrared divergences disappear
and the graphs can be safely computed at $m_W=0$\footnote{We have checked that the more cumbersome
procedure of keeping $m_W$ as an infrared regulator gives the same final result.
When doing the computation at $m_W=0$, it would be useful to know if the $1/\eps$ poles
are due to infrared (IR) or to ultraviolet (UV) divergences.
However some of the graphs in fig.~\ref{fig:gra2} contain terms like $\eps/\epsUV\epsIR$,
so that we have not been able
to distinguish IR from UV divergences using dimensional regularization only.}.

\item Finally we have to renormalize the CKM mixing angles.
Since only the mixing between $2^{\rm nd}$ and $3^{\rm rd}$ generation is relevant,
the mixing is described by only one parameter.
Furthermore we may set $V_{tb}=1$, $V_{ts}=-V_{cb}=s_{23}\ll 1$:
we can neglect the $\Ord(g_t,\lambda)^2$ corrections to
terms suppressed by powers of the Wolfenstein parameter $\lambda^2_{\rm W}\approx 0.05$~\cite{Wolfenstein}.
At this point, we remember that normalising the $b\to s\gamma$ decay rate to the semi-leptonic $b\to c \bar{e}\nu$ one
allows to eliminate a large hadronic uncertainty coming from a $m_b^5$ phase-space factor.
For this reason, we will extract the renormalized value of $s_{23}$
from  $\BR(B\to X_c \bar{e}\nu)$.
\end{itemize}
We now show how is it possible to accomplish the last point.
It is {\em very convenient} to renormalize a non standard form
of the bare Lagrangian that is respected by the quantum corrections that we are considering.
Keeping only the bare interactions and the bare fields necessary to describe
flavour mixing, we write
the $\SU(2)_L$-symmetric bare Lagrangian as
\begin{equation}\label{eq:L0}
{\cal L}_0=\sum_q \bar{q}^0i\Dsl q^0 + g_t^0 \, t_R^0 Q_3^0 H^{*0}+
g_b^0 \, b_R^0 ( Q_3^0 - s_{23}^0 Q_2^0)H^0
\end{equation}
where $q=\{Q_3^0,Q_2^0,t_R^0,b_R^0\}$.
The down components of the quark doublets, ${\rm Q}_{\downarrow}^0$, are not mass eigenstates.
Had we started from the usual flavour basis of mass eigenstates,
complicated flavour redefinitions would be necessary to obtain a unitary renormalized CKM matrix.
The tree level Lagrangian and  the divergent part of the quantum corrections have a $\SU(2)_L$ global symmetry.
This symmetry is useful to understand some cancellations (for example why $Z_{m_t}$ is not divergent)
and to fix some dangerous signs.
It is easy to verify that the 1PI effective Lagrangian, corrected {\em only}
by the divergent part of the quantum corrections is
$${\cal L}_{\rm effective}=\sum_q Z_q \bar{q}^0i\Dsl q^0 + Z_{g_t} g_t^0 \, t_R^0 Q_3^0 H^{*0}+
g_b^0 \, b_R^0 (Z_{g_b}   Q_3^0  - s_{23}^0 Q_2^0)H^0$$
and has the same form as the bare Lagrangian\eq{L0}
(its last term does not get $\Ord(g_t^2)$ corrections).
The $Z$ factors are
$$Z_{Q_3}=1+\frac{1}{2}\frac{g_t^2}{(4\pi)^2}\frac{1}{\eps},\qquad
Z_{t_R}=Z_{g_b}=1+\frac{g_t^2}{(4\pi)^2}\frac{1}{\eps},\qquad
Z_{Q_2}=Z_{g_t}=Z_{b_R}=1.$$
We are now ready to renormalize also the finite part of the quantum corrections.
Performing, in the effective Lagrangian, the following redefinitions of the fields
\begin{eqnsystem}{sys:rin}&&
t_R^0=\frac{t_R}{\sqrt{Z_{t_R}}},\qquad
b_R^0=\frac{b_R}{\sqrt{Z_{b_R}}},\qquad
Q_3^0={t_L/\sqrt{Z_{t_L}}\choose ( b_L+s_{23}\, s_L)/\sqrt{Z_{b_L}}},\qquad
Q_2^0={c_L \choose  s_L - s_{23}\, b_L}\\
\riga{and couplings}\\
&&
vg_t^0=m_t\frac{\sqrt{Z_{t_R}Z_{t_L}}}{Z_{m_t}},\qquad
vg_b^0=m_b \frac{\sqrt{Z_{b_L}}}{Z_{m_b}},\qquad
s_{23}^0=\frac{Z_{m_b}}{\sqrt{Z_{b_L}}}s_{23}
\end{eqnsystem}
the renormalized parameters $m_t$, $m_b$ and $s_{23}$ coincide with the
physical quantities previously chosen
(the meaning of the $Z$ factors should be clear from the notation;
they are precisely defined and computed in appendix~B).
Notice that, only at this stage, we have performed a unitary flavour rotation in the down
components of the quark doublets Q,
so that the renormalized quark fields are mass eigenstates.
The $V_{cb}$ CKM mixing angle that appears at the gauge $\bar{c}_L\Wsl b_L$ vertex is $s_{23}$
(the gauge interaction of $Q^0_2$ does not get $\Ord(g_t,\lambda)^2$ corrections), so that
the link between $B\to X_s\gamma$ and the semi-leptonic decay $B\to X_c\bar{e} \nu$ decay
can now be easily extracted.

\subsection{Renormalization of $C_7$}
We are now ready to express the one-loop $b\to s \gamma$ magnetic penguin
in terms of renormalized parameters and fields.
The magnetic penguin obtained from the (unusual) tree level Lagrangian\eq{L0} is
\begin{equation}
{\cal H}_{\rm top~penguins}^{\rm 1~loop}=-\left\{
\frac{5}{12}\frac{g_t^0 g_b^0}{m_t^0}  {\cal O}[b_R^0,Q_{3\downarrow}^0] -
\frac{1}{9} \frac{g_t^{02}}{m_t^{02}} {\cal O}[Q_{3\downarrow}^0,i\ds Q_{3\downarrow}^0]\right\}
\left(\frac{\mub}{m_t^0}\right)^{2\eps}
\end{equation}
where ${\cal O}[a,b]=\frac{e}{(4\pi)^2}(\bar{a}\gamma_{\mu\nu}  \pR b)F_{\mu\nu}$.
Notice that no explicit mixing angle appears at this stage.
The two contributions arise respectively from the one-loop graphs named ``I'' and ``E'' in fig.~\ref{fig:gra1}
and give rise to two different operators, that are equivalent only on-shell.
Since the graph ``E'' is not one-particle irreducible, we must take
this fact into account in the renormalization.
Operating the redefinitions\eq{MW<-->v} and~(\ref{sys:rin}), and using
the renormalized Dirac equation $i\ds b=m_b b$ to set the $b$-quark on-shell, we obtain
\begin{equation}\label{eq:rin}
{\cal H}_{\rm top~penguins}^{\rm 1~loop}=
-\frac{\GF}{\sqrt{2}}s_{23}\left\{\frac{5}{12}
\frac{Z_\chi}{Z_{m_b}}-\frac{1}{9}\frac{Z_\chi}{Z_{b_L}} \right\}
\big[1+\eps\ln\frac{Z_{m_t}^2}{Z_{t_L}Z_{t_R}}+\Ord(\eps^2)\big]  \Op{7}.
\end{equation}
The final formula for the coefficient $C_7$ of the
$b\to s\gamma$ magnetic penguin, that includes all
renormalized $\Ord(g_t,\lambda)^2$ corrections, is thus
\begin{equation}
\scatola{\Black C_7=-\frac{23}{36}+
\frac{5}{12}\frac{Z_\chi}{Z_{m_b}}-
\frac{1}{9}\frac{Z_\chi}{Z_{b_L}}-\frac{g_t^2}{(4\pi)^2}\frac{11}{24}+
{\hbox{two-loop diagrams}\choose
\hbox{plotted in figure~\ref{fig:gra2}}}.\Blue }
\end{equation}
We have added the charm contribution, $-23/36$, that does not get $\Ord(g_t,\lambda)^2$ corrections.
In conclusion the heavy top and/or heavy higgs correction to $C_7$, computed in the limit $m_W\to 0$, is
$$C_7(m_t^2\gg m_W^2)=-\frac{1}{3}+\frac{g_t^2}{(4\pi)^2}C_7^{g_t^2}$$
with
\begin{eqnarray*}
C_7^{g_t^2}&=&
\frac{-16 + 39r - 11{r^2} - 26{r^3}}{144r} - 
\frac{-16 + 2r - 36{r^2} + 74{r^4} - 45{r^5} + 2{r^6} }{864 r^2} \pi^2+ \\
&&-
\frac{8 - r - 6{r^2} - 52{r^3} + 85{r^4} - 33{r^5} + 2{r^6} }{72{r^2}} \Li(1 - r) -
\frac{80 + 68r - 262{r^2} + 134{r^3} - 25{r^4} + 2{r^5} }{288r}\Phi (\frac{r}{4})+\\
&&+
\frac{8  - 17r - 2 r^2 - 14{r^3} }{72r}\ln r -
\frac{74 - 45r + 2 r^2}{288} r^2 \ln^2 r \approx  0.14-0.0024 r-0.046\ln r
\end{eqnarray*}
where $r=m_h^2/m_t^2$, the functions $\Li$ and $\Phi$ are defined in appendix~\ref{f},
and the approximation holds for any reasonable value of $m_h\in[60,1000]\GeV$.
Since, instead of using $\MSbar$ subtractions,
we have expressed the result in terms of physical quantities,
no dependence on the $\MSbar$ scale $\mub$ is left.
In the limits $m_h\ll m_t$ and $m_h\gg m_t$ $C_7$ reduces to
$$C_7^{g_t^2}(r\to 0)=N_c(\frac{\pi^2}{36}-\frac{1}{9})-
(\frac{\pi^2}{36}-\frac{1}{16}),\qquad
C_7^{g_t^2}(r\to \infty)=-\frac{19}{144}\ln r +\frac{17}{432}+\frac{\pi^2}{24}$$
where $N_c=3$ is the number of colours.
In table~\ref{tab:lims} the asymptotic dependences of these corrections in the limits of heavy top and of heavy higgs
are compared to the corresponding limits of the corrections to the observables measured at LEP.

\begin{figure}[t]\setlength{\unitlength}{1in}\begin{center}
\begin{picture}(5,3)
\ifMac
{\put(0,-0.3){\special{picture res}}}
{\put(0,-0.3){\includegraphics{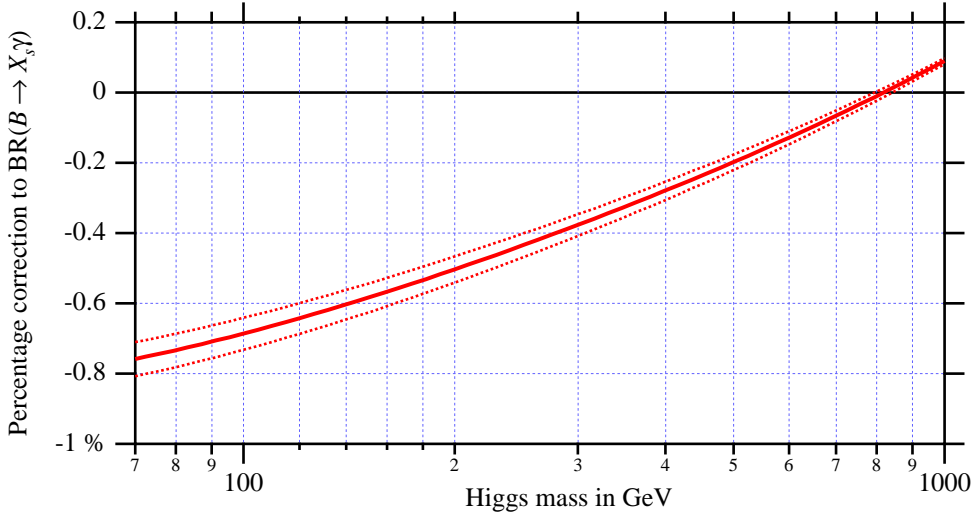}}}
\end{picture}
\vspace{0.5cm}
\caption[SP]{\em The percentage correction to $\BR(B\to X_s\gamma)$ for $m_t({\rm pole})=(175\pm 5)\GeV$
and as function of the Higgs mass
in the limit of heavy top ad/or higgs.
\label{fig:res}}
\end{center}\end{figure}

\subsection{Branching ratio}
With our renormalization scheme the $B\to X_s \gamma$ branching ratio can be linked to the
$b\to s\gamma$ decay width in the usual way (see~\cite{NLO7} and references therein),
that we now briefly recall.
Neglecting, for simplicity, QCD corrections
\begin{eqnsystem}{sys:BR/BR}
\BR(B\to X_s\gamma) &=&\frac{\Gamma(b\to s\gamma)}{\Gamma(b\to c\bar{e}\nu)}\BR(B\to X_c\bar{e}\nu)\\
\riga{where $\BR(B\to X_c\bar{e}\nu)=0.105\pm 0.05$ has been measured~\cite{BRSL},}\\[-3mm]
\Gamma(b\to s\gamma) &=&\frac{G_{\rm F}^2}{32\pi^3} m_b^5 (s_{23} c_{23})^2 \frac{e^2}{4\pi^2}|C_7(\mub_b)|^2,\\
\Gamma(b\to c\bar{e}\nu)&=&\frac{G_{\rm F}^2}{32\pi^3} m_b^5 (s_{23})^2 \frac{g(z)}{6},\qquad
z\equiv\frac{m_c^2}{m_b^2}
\end{eqnsystem}
and $g(z)$ is a phase space factor~\cite{NLO7}.
The $\Ord(g_t,\lambda)^2$ corrections to the the semileptonic width are entirely
contained in our definitions of the renormalized parameters.
As previously said the complete CKM mixing can be reinserted
neglecting the $\Ord(g_t,\lambda)^2$ corrections to the terms suppressed by powers of $\lambda_W^2$:
$(s_{23}c_{23})^2/(s_{23})^2\to |V_{tb}V_{ts}/V_{cb}|^2 = 1+\lambda_{\rm W}^2(2\rho_{\rm W}-1)+\Ord(\lambda_{\rm W}^4)$,
($\lambda_{\rm W}\approx 0.22$ and $\rho_{\rm W}$ are Wolfenstein parameters~\cite{Wolfenstein}).

These expressions receive important {\em QCD corrections} (perturbative and non perturbative), that
can be added in the usual way.
Infact, the $\Ord(g_t,\lambda)^2$ corrections that we are considering,
only affect the values of the Wilson coefficients at the
electroweak scale, $C_i(\mub_W)$, $i=1,\cdots,8$.
We have computed the correction to $C_7$.
Other relevant $\Ord(g_t,\lambda)^2$ corrections can be present in $C_2$ (the coefficient of the
$(\bar{s}^i\gamma_\mu {\cal P}_L c^i)(\bar{c}^j\gamma_\mu {\cal P}_L b^j)$ operator)
and in $C_8$ (the coefficient of the $b\to sg$ chromo-magnetic penguin).
In a formal expansion in powers of $\alpha_3$,
the corrections to $C_2$ and $C_8$ enter at the same order as the correction to $C_7$.
\begin{itemize}
\item However, the correction to $C_8$ cannot have a significant effect on the $B\to X_s \gamma$ decay,
in view of the small mixing coefficient between $C_7$ and $C_8$:
$$C_7(\mub_b)\approx -0.155+0.65\, C_7(\mub_W)+0.085\, C_8(\mub_W),$$
where $\mub_b\sim m_b$ and $\mub_W\sim m_W$.
In any case, it is immediate to obtain the renormalized correction to $C_8$
from a subset of the two-loop graphs of fig.~\ref{fig:gra2}.
In the heavy top limit $C_8$
has a weak dependence on the higgs mass $m_h$
and can be approximated with its value at $m_h=0$,
$C_8(\mub_W)=-1/8+g_t^2/(4\pi)^2 (\pi^2/12-13/32)$.

\item The $\Ord(g_t,\lambda)^2$ correction to $C_2$ vanishes in our renormalization scheme.
\end{itemize}
We remember that the coefficient $C_7$ can also be extracted
from the spectrum of the $B\to X_s\ell^+\ell^-$ decays~\cite{generalSUSYinB}.

\section{Numerical result}
Due to a numerical accident, the potentially relevant heavy top correction to
the $b\to s\gamma$ magnetic penguin turns out to be negligible,
if the leading order result is expressed in terms of the renormalized parameters
$\GF$ and $\BR(B\to X_c \bar{e}\nu)$.
The percentage correction to $\BR(B\to X_s\gamma)$ is plotted in fig.~\ref{fig:res}
as function of the higgs mass, $m_h$, and for $m_t=(175\pm 5)\GeV$,
where $m_t$ is the pole top mass.
In the plot we have included the QCD corrections at NLO order and
all the heavy top and/or heavy higgs corrections
(also the small ones, not fully given in the text, due to the mixing with the chromo-magnetic penguin).

We see that the correction is never larger than $1\%$ for any possibly interesting value of the higgs mass.
We remember that small values of the Higgs mass are preferred.
The electroweak precision measurements give $m_h<250\GeV$ at $90\%$ C.L.~\cite{EWdata}.
Moreover, the minimal supersymmetric extension of the SM (MSSM) predicts a light higgs:
at tree level it requires that $m_h< m_Z$.
Adding the one-loop corrections~\cite{mh1loop} and
barring an unnatural fine-tuning larger than 20,
the MSSM bound on the higgs mass becomes $m_h<120\GeV$~\cite{mhFT}.

As said, the smallness of the correction is due to a numerical accident.
Before having done this computation a relevant heavy top correction,
at the $5\%$ level or even larger\footnote{Even if $g_3(\mub_W)\approx 1.2$ is not significantly larger than $g_t\approx 1.0$,
the well known NLO QCD effects
of relative order $g_3^2/(4\pi)^2$ give significantly larger corrections (at the $20\%$ level)
because they have `more colour' and `more spin' circulating in the single Feynman graphs.},
could not be excluded.
However, due to accidental cancellations between different contributions these corrections
are smaller than $1\%$.

\section{Conclusions}
We have computed the heavy top and the heavy higgs corrections to
the Wilson coefficient of the $b\to s\gamma$ magnetic penguin in the limit $m_W\to 0$.
These potentially relevant corrections turn out to be very small.
Since there are accidental cancellations between different contributions,
it is possible that the limit $m_W\to 0$ be not a good approximation.
However it appears unlikely that the neglected corrections,
suppressed by powers of $m^2_W/m^2_t$, be sufficiently large to
make the heavy top effects relevant\footnote{The ones enhanced by
a $N_c$ factor have recently been computed in~\cite{CzMa}.
Including QCD corrections they give another $-1\%$ correction to $\BR(B\to X_s\gamma)/\BR(B\to X_c\bar{e}\nu)$.}
For these reasons it seems safe to conclude that these kind of corrections can be neglected.

\paragraph{Acknowledgements}
I thank R. Barbieri for several
illuminating discussions and M. Ciuchini, G. Degrassi and G. Giudice,
that collaborated in an early stage of this work.

\appendix

\setcounter{equation}{0}
\renewcommand{\theequation}{\thesection.\arabic{equation}}

\section{The relevant Lagrangian}
As explained in the text we can work in the gauge-less limit.
In a flavour basis
where the Yukawa matrices $g_{\rm u}$ and $g_{\rm d}$ are diagonal,
the relevant Lagrangian is
\begin{equation}
{\cal L} = \bar{\rm q}i\ds {\rm q} + |\partial H|^2 + \mu^2 |H|^2 - \lambda |H|^4 -
 (\bar{\rm u}_R g_{\rm u} {\rm Q}  H^*+
  \bar{\rm d}_R g_{\rm d}V^\dagger{\rm Q}  H +\hbox{h.c.})
\end{equation}
where $V$ is the CKM matrix, $\hbox{d}=\{d,s,b\}$, $\hbox{u}=\{u,c,t\}$, $\rm q=\{u,d\}$.
The contractions of the $\SU(2)_L$  doublets ${\rm Q}$ and $H$, expressed in terms of their
up ($\uparrow$) and down ($\downarrow$) components, are
$$\bar{\rm u}_R QH^*=\bar{\rm u}_R({\rm Q}_\uparrow H_\uparrow^*+ {\rm Q}_\downarrow H_\downarrow^*)\qquad\hbox{and}\qquad
\bar{\rm d}_R QH=\bar{\rm d}_R({\rm Q}_\downarrow H_\uparrow - {\rm Q}_\uparrow H_\downarrow)$$
so that the mass terms of the quarks have conventional sign.
Decomposing the fields into mass eigenstates
(we denote the unphysical higgses with greek characters)
\begin{equation}
H=\pmatrix{v+(h+i\eta)/\sqrt{2} \cr  \chi_-},\qquad
\rm{Q} = \pmatrix{{\rm u}_L \cr V {\rm d}_L}
\end{equation}
one obtains
\begin{equation}\label{eq:Lclass}
{\cal L} = {\cal L}_{\rm free} +{\cal L}_{\rm Yuk} -
V_{\rm cubic}-V_{\rm quartic}+\sqrt{2}v(\mu^2-2\lambda v^2) h
\end{equation}
where
\begin{eqnarray*}
{\cal L}_{\rm free} &=& \bar{\rm q}(i\ds-m_{\rm q}) {\rm q}+
|\partial \chi|^2 + \frac{(\partial h)^2}{2} +\frac{(\partial\eta)^2}{2} -
( 6\lambda v^2-\mu^2)\frac{h^2}{2}+(\mu^2-2\lambda v^2)(\frac{\eta^2}{2}+|\chi|^2)\\
 {\cal L}_{\rm Yuk} &=& -g_t\left[\bar{t}\,\frac{h+i\gamma_5 \eta}{\sqrt{2}} t +
 (\chi_- ~V_{t\rm d} \bar{\rm d}_L t_R+\hbox{h.c.})\right]
-g_b\left[\bar{b}\,\frac{h-i\gamma_5 \eta}{\sqrt{2}} b -
 (\chi_- ~V_{tb}\bar{b}_R {t}_L+\hbox{h.c.})
\right]\\
V_{\rm cubic}&=&\sqrt{2}\lambda v (h^3 + h\eta^2 + 2 h \chi_-\chi_+)\\
V_{\rm quartic}&=&\lambda \chi_-\chi_+(h^2+\eta^2+\chi_-\chi_-)+\frac{\lambda}{4}(h^2+\eta^2)^2.
\end{eqnarray*}

\section{Renormalization constants}
In this appendix we give the $Z$ constants needed for renormalization.

\subsection{Bottom quark}
We write the inverse 1-loop propagator of a $b$ quark with momentum $p_b$ as
$\pl_b [Z_{b_L} {\cal P}_L+Z_{b_R} {\cal P}_R]-Z_{m_b} m_b $
where ${\cal P}_{L,R}=(1\mp\gamma_5)/2$.
Since there are no infrared divergences even in the limit $m_W=0$,
the constants $Z$ depend on $p_b$ only trough irrelevant terms suppressed by powers of
$p_b^2/m_t^2$. Their values are: $Z_{b_R}=1$,
$$
Z_{b_L}=1+\frac{g_t^2}{(4\pi)^2}\left[
\frac{1}{2}(\frac{1}{\varepsilon}+\ln\frac{\mub^2}{m_t^2})+\frac{3}{4}\right]\qquad\hbox{and}\qquad
Z_{m_b}=1+\frac{g_t^2}{(4\pi)^2}\left[\frac{1}{\varepsilon}+\ln\frac{\mub^2}{m_t^2}+1\right].
$$

\subsection{Top quark}
We write the inverse 1-loop propagator of a $t$ quark with momentum $p_t$ as
$\pl_t [Z_{t_L} {\cal P}_L+Z_{t_R} {\cal P}_R]-Z_{m_t} m_t $.
We need only the divergent parts of the $Z$ constants.
These divergent parts do not depend on $p_t$ and can be more easily
computed in the limit of unbroken $\SU(2)_L$ global symmetry.
We find:
$$
Z_{m_t}=1+\frac{g_t^2}{(4\pi)^2}\left[\frac{0}{\eps}+\cdots\right],\qquad
Z_{t_R}=1+\frac{g_t^2}{(4\pi)^2}\left[\frac{1}{\eps}+\cdots\right],\qquad
Z_{t_L}=1+\frac{g_t^2}{(4\pi)^2}\left[\frac{1}{2\eps}+\cdots\right]
$$
where $\cdots$ denotes the finite parts that we do not need.

\subsection{Higgs}
We write the one-particle irreducible (1PI)
propagator of a unphysical charged higgs $\chi$ with momentum $p_\chi$ as
$i/[q^2 Z_\chi(p_\chi^2)  -  m_\chi^2 Z_{m_\chi}(p_\chi^2)]$.
As explained in the text we enforce the vanishing of Goldstone masses, $m_\chi^2=0$,
inserting appropriate counterterms
(equivalent to the non 1PI tadpole graphs that shift the vev $v$ to the quantum corrected minimum).
We only need the constant $Z_\chi$ at zero momentum.
At one loop\footnote{The corresponding expression given by Barbieri et.\ al in~\cite{mt^4}
contains a misprint (not present in their final result).}
\begin{equation}
Z_\chi=1+\frac{1}{(4\pi)^2}\left[\lambda + g_t^2 N_c
(\frac{1}{\varepsilon}+\lnEps + \frac{1}{2})\right]
\end{equation}
with $\lambda=+g_t^2 r/4$.

\section{Special functions}\label{f}
Our final result contains the bi-logarithmic function
$\Li(z)$ defined as 
$\Li(x)\equiv -\int_0^1 \ln (1-xt)\,dt/t$
and the function
$$\Phi(z)\equiv\left\{\begin{array}{ll}
\sqrt{\frac{z}{1-z}}\, 4\, \hbox{Im} \,\Li \exp{(2i\arcsin\sqrt{z})} & \hbox{for $0<z<1$}\\
\sqrt{\frac{z}{z-1}}\left[\frac{\pi^2}{3}-\ln^2(4z)+2\ln^2\frac{1-\sqrt{1-1/z}}{2}-4
\Li\frac{1-\sqrt{1-1/z}}{2}\right]&
\hbox{for $z>1$}
\end{array}\right.$$
The one loop penguin functions used in\eq{P} are
\begin{eqnsystem}{sys:P}
P_W(x)&=&\frac{-23 + 67\,x - 50\,{x^2}}{36(x-1)^3} + 
\frac{2 - 7\,x + 6\,{x^2}}{6(x-1)^4}x\ln x\\
P_I(x) &=&\frac{3 - 5\,x}{6(x-1)^2} + \frac{3x-2}{3(x-1)^3}\ln x\\
P_E(x)&=&\frac{7 - 5\,x - 8\,{x^2}}{36(x-1)^3} + \frac{3x-2}{6(x-1)^4}x\ln x
\end{eqnsystem}

\end{document}
\\
Title: Two-loop heavy top corrections to the $b\to s \gamma$ decay
Authors: Alessandro Strumia
Comments: 10 pages.
Report-no: IFUP-TH 16/98
\\
We compute the two-loop corrections to the coefficient
of the b --> s gamma magnetic penguin present in the limits of
heavy top and/or heavy higgs.
This kind of corrections affects in a significant way the observables measured at LEP.
On the contrary we find that, due to a numerical accident,
the correction to BR(B --> X_s gamma) is negligible (below the 1
for any possible value of the higgs mass)
when the leading order result is expressed in the usual way
in terms of the semileptonic BR(B --> X_c \bar{e}\nu).
\\